\documentclass[%
 reprint,
 amsmath,amssymb,
 aps,
]{revtex4-2}
\usepackage{diagbox}
\usepackage{graphicx}
\usepackage{dcolumn}
\usepackage{bm}

\usepackage{gensymb}
\usepackage{hyperref}
\usepackage{amsmath}
\usepackage{subcaption}
\usepackage{booktabs}
\usepackage{braket}
\usepackage{stmaryrd}
\usepackage{multirow}
\usepackage{subcaption}
\usepackage{xcolor}
\usepackage{makecell}
\usepackage{verbatim}
\usepackage{etoolbox}

\begin{document}

\preprint{APS/123-QED}

\title{The correlated matching decoder for the 4.8.8 color code}

\author{Yantong Liu}
\affiliation{%
College of Computer Science and Technology, National University of Defense Technology, Changsha 410073, China
}

\author{Junjie Wu}
 \email{junjiewu@nudt.edu.cn}

\affiliation{%
College of Computer Science and Technology, National University of Defense Technology, Changsha 410073, China
}

\author{Lingling Lao}%
 \email{laolinglingrolls@gmail.com}
\affiliation{%
College of Computer Science and Technology, National University of Defense Technology, Changsha 410073, China
}
\date{\today}

\begin{abstract}
Color codes present distinct advantages for fault-tolerant quantum computing, such as high encoding rates and the transversal implementation of Clifford gates.
However, existing matching-based decoders for the color codes such as the restricted decoder (Kubica and Delfosse, 2023), suffer from limited decoding performance. Inspired by the global decoding insight of the unified decoder (Benhemou et al., 2023), this paper introduces a correlated decoder for the 4.8.8 color code, which improves upon the conventional restricted decoder by leveraging correlations between restricted lattices, and is derived by mapping the correlated matching decoder for the surface code onto the color code lattice.
Analytical and numerical results show that the correlated decoder achieves  higher thresholds than the restricted and unified decoders, while matching the performance of the unified decoder at very low physical error rates.
Under the code capacity and phenomenological noise models, the estimated thresholds for the color code against bit-flip error are 10.38\% and 3.13\%, respectively. Furthermore, by applying the surface-color code mapping, the thresholds of 16.62\% and 3.52\% are obtained for the surface code against depolarizing noise. 

\end{abstract}

\maketitle


\section{\label{sec:introduction}Introduction}

Quantum computing promises substantial speed-ups over 
classical computing for specific problems. However, the practical realization of quantum algorithms is hindered by environmental noise and imperfect control, which can rapidly corrupt the fragile quantum states of qubits. Quantum error correction (QEC) provides a pathway towards fault tolerance by encoding logical quantum information into entangled states of many physical qubits, thereby protecting it from local errors~\cite{shor1995scheme,shor1996fault,calderbank1996good,terhal2015quantum}. A QEC protocol involves repeatedly measuring stabilizers to extract \textit{syndrome} information, which is then processed by a \textit{decoder} to infer the most likely error pattern. Based on this inference, errors can either be actively corrected or accounted for in subsequent logical measurements.
The quantum threshold theorem~\cite{aharonov1997fault,preskill1998reliable,shor1996fault} states that the logical error rate can be suppressed exponentially by increasing the code distances, provided the physical error rate lies below a critical value known as the \textit{threshold}. The threshold depends strongly on the choice of QEC codes, the noise model, and the decoding algorithm, making the development of high-performance and low-overhead decoders essential for scalable quantum computation.

Topological codes, particularly surface codes~\cite{kitaev2003fault, bravyi1998quantum, dennis2002topological, Fowler2012surface} and color codes~\cite{bombin2006topological, bombin2007topological, bombin2015gauge, Kubica2015universal, kubica2018abcs}, are leading candidates for practical fault-tolerant quantum computing. Their appeal stems from locality of stabilizer checks, compatibility with near-term hardware architectures, and high error thresholds. Compared to surface codes, color codes demonstrate several noteworthy advantages, such as achieving higher encoding rates for a given code distance~\cite{landahl2011fault}, and permitting transversal implementation of all Clifford gates~\cite{bombin2006topological}. To fully exploit these properties in fault-tolerant architectures, high-performance decoding algorithms for color codes are indispensable.

A variety of decoders have been developed for color codes. A prominent and efficient class is based on the minimum-weight perfect-matching (MWPM) algorithm~\cite{edmonds1965paths,wu2023fusion,higgott2025sparse, higgott2022pymatching}, which pairs syndrome excitations on a matching graph. For instance, the projection decoder and its variants~\cite{delfosse2014decoding_cc, chamberland2020triangular, kubica2023efficient} decode by applying MWPM independently on two or three restricted lattices---sublattices derived by removing one color of checks. However, ignoring correlations between these sublattices makes such decoders exhibit a fundamental limitation: they fail to decode certain error configurations of weight-$O(d/3)$ on the 6.6.6 color code~\cite{delfosse2014decoding_cc, sahay2022decoder} and weight-$O(d/2)$ on the 4.8.8 color code~\cite{benhemou2025minimising}. Since the logical failure rate in the low-error regime is dominated by minimum-weight failure events, this limitation directly impacts decoding performance.
 The M{\"o}bius decoder~\cite{sahay2022decoder} for 6.6.6 color code and its variants---the unified decoder~\cite{benhemou2025minimising} for 4.8.8 color code both decode by applying MWPM on the manifold connected by the three restricted lattices 
 . They resolve the decoding problems aforementioned, thereby achieving a better performance. The concatenated MWPM decoder also exploits the correlation of restricted lattices to decode~\cite{lee2025color}. Nevertheless, these decoders have a higher matching overhead due to their more complex matching graphs. In comparison, our correlated decoder effectively leverages restricted lattice correlations to resolve the same problem while preserving the low computational overhead of the restricted decoder.
 
 Besides these matching-based decoders, a wide range of other decoders have been explored, whose representative thresholds of decoding color codes under bit-flip noise are summarized in Table~\ref{tab:thresholds}. Furthermore, for practical quantum computing, decoders must handle phenomenological and circuit-level noise models. Significant progress has been made under this consideration as well~\cite{gidney2023new, lee2025color, koutsioumpas2025colour, landahl2011fault, wu2025minimum, beni2025tesseract}.

\begin{table}[h]
    \centering
    \begin{tabular}{|c|c|c|}
    \hline
    Lattice & Decoder & Threshold \\
    \hline
    \multirow{9}{*}{\makecell{Hexagon\\(6.6.6)}} & Optimal & 10.9\%~\cite{katzgraber2009error} \\
           & Tensor Network & 10.9\%~\cite{chubb2021general} \\
           & Neural Network & 10.0\%~\cite{maskara2019advantages} \\
           & M{\"o}bius MWPM & 9.0\%~\cite{sahay2022decoder} \\
           & Restricted MWPM & 8.7\%~\cite{delfosse2014decoding_cc} \\
           & Concatenated MWPM & 8.2\%~\cite{lee2025color} \\
           & Union Find & 8.4\%~\cite{delfosse2021almost} \\
           & Renormalization-Group & 7.8\%~\cite{sarvepalli2012efficient} \\

    \hline
    \multirow{9}{*}{\makecell{Square-Octagon\\(4.8.8)}} & Optimal & 10.9\%~\cite{katzgraber2009error} \\
        & Integer program & 10.5\%\cite{landahl2011fault} \\
        & $\textbf{Correlated MWPM}$ & \textbf{10.38\%} \\
        & Ising model & 10.3\%~\cite{takada2024ising} \\
        & Restricted MWPM & 10.2\%~\cite{kubica2023efficient} \\
        & Unified MWPM & 10.1\%~\cite{benhemou2025minimising} \\
        & Union-Find & 9.8\%~\cite{kubica2023efficient} \\
        & Renormalization-Group & 8.7\%~\cite{bombin2012universal} \\
        
    \hline
    \end{tabular}
    \caption{Thresholds of different decoders for the 6.6.6 and 4.8.8 color code lattices against bit-flip error.}
    \label{tab:thresholds}
\end{table}

In this work, we propose a correlated decoder for the 4.8.8 color code, which considers the correlations between restricted lattices for decoding. Through the color-surface code mapping~\cite{yoshida2011classification, bombin2012universal, Bhagoji15, Kubica_2015, criger2016noise, benhemou2025minimising}, we illustrate the correlated (restricted) decoder for the color code are the natural analogues of the correlated (independent) $X$/$Z$ MWPM decoder~\cite{delfosse2014decoding_sc, fowler2013optimal} for surface code. We analyze the decoding performance for the minimum-weight ($O(d/2)$) uncorrectable error chains, demonstrating its exponential improvements over the restricted decoder at very low physical error rates. We further numerically estimate the thresholds for the restricted, the unified and our correlated decoders, confirming the advantages of the correlated decoder in both error thresholds and logical error rates.  

The remainder of this article is organized as follows. Sec.~\ref{sec:preliminaries} reviews stabilizer codes and the decoding problem. Sec.~\ref{sec:Mapping} details the mapping from surface code to color code which is the key to understanding this article. We present our correlated decoder in Sec.~\ref{sec:correlated decoder} and evaluate its decoding performance in Sec.~\ref{sec:performance analysis} and Sec.~\ref{sec:results}. Sec.~\ref{sec:discussion} concludes the paper.

\section{\label{sec:preliminaries}Preliminaries}

\subsection{Stabilizer codes}
Stabilizer codes~\cite{gottesman1998theory} constitute the most prominent and widely used family of quantum error-correcting codes, including topological codes such as the surface code and the color code. 
The single-qubit Pauli group is defined as $\mathcal{P}=\{\pm1,\pm i\}\times\{I,X,Y,Z\}$, consisting of the Pauli operators with phases. 
The $n$-qubit Pauli group $\mathcal{P}_n$ is given by the $n$-fold tensor product of $\mathcal{P}$. An $n$-qubit stabilizer code is defined by an Abelian subgroup of Pauli group $\mathcal{S}\subset \mathcal{P}_n$, known as the stabilizer group, that does not contain the element $-I$. Suppose $\mathcal{S}$ has $n-k$ independent generators. The code space $V_\mathcal{S}$ is the $2^k$-dimensional subspace of the Hilbert space spanned by all states that are simultaneous $+1$-eigenvectors of every stabilizer in $\mathcal{S}$:
\begin{equation}
    V_\mathcal{S}=\text{span}\{\vert\psi_L\rangle|s|\psi_L\rangle=+1|\psi_L\rangle,\forall s\in \mathcal{S}\}
\end{equation}
The logical operators of the code are given by the set $\mathcal{L} = C(\mathcal{S}) \setminus \mathcal{S}$, where $C(\mathcal{S})$ denotes the centralizer of $\mathcal{S}$ in Pauli group. A stabilizer code is characterized by the parameters $\llbracket n,k,d \rrbracket$, where $n$ is the number of physical qubits, $k$ is the number of logical qubits, and the code distance $d$ is defined as the minimum weight among all non-trivial logical operators in $\mathcal{L}$.

Stabilizer codes facilitate error detection without disturbing the encoded logical state through the measurement of stabilizer generators (often termed \textit{checks}). 
A check yields a syndrome of $-1$ if it anti-commutes with the error $E$, and $+1$ otherwise.
A decoder then uses the syndrome information to infer a possible error $E'$, as detailed in Section~\ref{subsec:decoding}. The decoding is successful if $E'\cdot E\in \mathcal{S}$. A correction operator $C$ satisfying $C\cdot E' \in \mathcal{S}$ can subsequently be applied to restore the state to the code space. However, if the product $E' \cdot E$ is a non-trivial logical operator, a logical failure occurs.

\subsection{The surface code}
The unrotated surface code can be conceptually derived by introducing boundaries into the toric code~\cite{kitaev2003fault}. It is defined on a two-dimensional lattice with specific boundary conditions, as illustrated in Fig.~\ref{fig:mapping_sc_cc}(c). Physical qubits are placed on the edges of the lattice. Stabilizer generators consist of $X$-type operators $S^X_v$ associated with each vertex $v$, and $Z$-type operators $S^Z_f$ associated with each face $f$. These are defined as the product of Pauli operators on the adjacent edges: 
$$S^X_v=\prod_{e\in E_v}X_e,\quad S^Z_f=\prod_{e\in E_f}Z_e$$
where $E_v$ and $E_f$ denote the sets of edges incident to vertex $v$ and surrounding face $f$, respectively. Smooth and rough boundaries are constructed by taking the product of $Z$ and $X$ operators along the top/bottom and left/right edges, respectively. The surface code encodes $k=1$ logical qubit with code distance $d$ using $n=2d^2-2d+1$ physical qubits.

\subsection{The color code}
The color code is defined on a two-dimensional lattice that is three-colorable. In this work, we focus specifically on the 4.8.8 color code, which is realized on a square-octagon lattice with a square boundary, as depicted in Fig.~\ref{fig:mapping_sc_cc}(a). Physical qubits are placed on the vertices of the lattice. Both $X$-type and $Z$-type stabilizer generators, denoted $S^X_f$ and $S^Z_f$ respectively, are associated with every face $f$ of the lattice. They are defined as the product of Pauli operators over all qubits on the boundary $\partial f$ of the face:
\begin{equation}
    S^X_f=\prod_{v\in \partial f}X_v, \quad S^Z_f=\prod_{v\in \partial f}Z_v.
\end{equation}

Let $\mathcal{C} = \{\textbf{r}, \textbf{g}, \textbf{b}\}$ be a set of color labels. A proper coloring assigns a color $\textbf{u} \in \mathcal{C}$ to each face such that adjacent faces have distinct colors. It is also useful to assign colors to the boundaries and corners of the lattice. A boundary is assigned a color $\textbf{u} \in \mathcal{C}$ if no qubit lying on it belongs to a face of color $\textbf{u}$. A corner is assigned color $\textbf{u}$ if its vertex is incident to only one face of that color. A $\textbf{u}$-colored corner occurs at the intersection of two boundaries with colors $\textbf{v}$ and $\textbf{w}$, where $\textbf{u}\neq \textbf{v} \neq \textbf{w} \neq \textbf{u}$. The color code features two green boundaries, two blue boundaries, and four red corners, encoding $k=2$ logical qubits with an even code distance $d$ using $n=2(d-1)^2+2$ physical qubits. Representatives of the logical operators can be defined as:
\begin{equation}
    \overline{X}_{\textbf{u}} = \prod\limits_{v\in\delta\textbf{u}} X_v, \quad \overline{Z}_{\textbf{u}} = \prod\limits_{v\in\delta\textbf{u}} Z_v,
    \label{eq:logical_ops}
\end{equation}
where the product is taken over all vertices $v$ on the boundary $\delta\textbf{u}$ of color $\textbf{u} \in \{\textbf{g}, \textbf{b}\}$.

\subsection{\label{sec:Mapping}Mapping from surface code to color code}
\begin{figure*}[htb]
    \centering
    \includegraphics[width=0.99\linewidth]{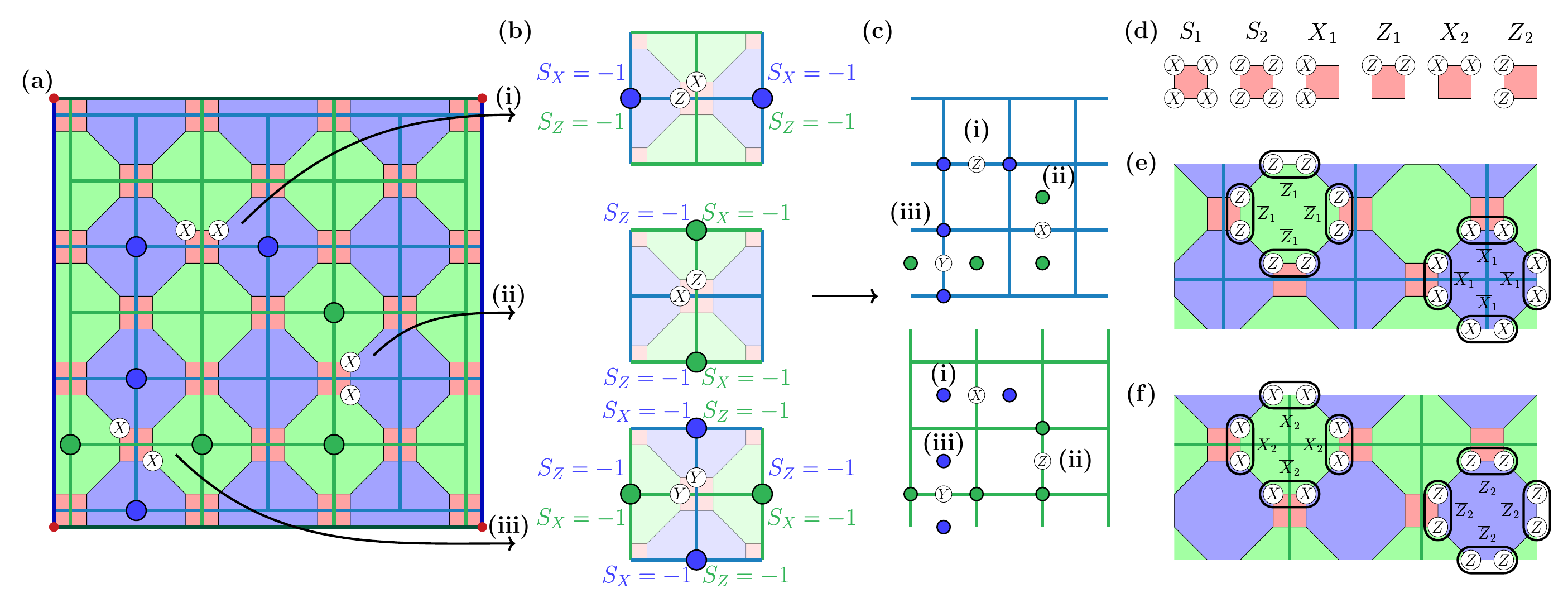}
    \caption{The surface-color code mapping. (a) The color code can be obtained by encoding physical qubits of a pair of surface codes (blue and green) using $\llbracket 4,2,2 \rrbracket$ codes. (b) Depolarizing errors on the blue and green surface codes are mapped to the two-qubit bit-flip errors on the color code (i.e., logical $X$ errors on the $\llbracket 4,2,2 \rrbracket$ codes). We emphasize that $\overline{X}_1\equiv \overline{Z}_2$ and $\overline{Z}_1\equiv \overline{X}_2$ with a Hadamard rotation under the surface code duplication. (c) The single-qubit depolarizing errors with their syndromes on the surface codes. (d) Stabilizers and logical operators of the $\llbracket 4,2,2 \rrbracket$ code defined on a square with qubits on its vertices. (e,f) For blue surface code (e), each physical qubit is encoded into the first logical qubit of a $\llbracket 4,2,2 \rrbracket$ code. The $X$ ($Z$) checks of the blue surface code are mapped to blue (green) octagonal $X$ ($Z$) checks of the color code. A similar mapping holds for the green surface code (f) after a transversal Hadamard rotation.}
    
    \label{fig:mapping_sc_cc}
\end{figure*}

This section describes the mapping from the surface code to the color code. We refer the readers to Ref.~\cite{benhemou2025minimising} for a detailed illustration. Through the mapping we can adopt color code decoders against bit-flip errors to the surface code against depolarizing errors. Furthermore, this mapping indicates that the restricted and correlated decoders for the color code can be derived from their counterparts for the surface code, as detailed in Sec.~\ref{sec:correlated decoder}.

The color code can be viewed as equivalent to two copies of the surface code~\cite{yoshida2011classification, bombin2012universal, Bhagoji15, Kubica_2015, criger2016noise, benhemou2025minimising}. The key insight is to construct the color code by encoding the physical qubits of a pair of overlapped surface codes (blue and green) using $\llbracket 4,2,2 \rrbracket$ codes (Fig.~\ref{fig:mapping_sc_cc}(a)). Each $\llbracket 4,2,2 \rrbracket$ code associated with the red square checks of the color code, encodes two corresponding physical data qubits of the two surface codes ( Fig.~\ref{fig:mapping_sc_cc}(c)). Consequently, the checks of the two surfaces codes are mapped to octagonal checks of the color code, see Fig.~\ref{fig:mapping_sc_cc}(d,e). 

Let's consider the mapping of the blue surface code as an example (the green one follows analogously). Each physical qubit of the blue code is encoded into the first logical qubit of a $\llbracket 4,2,2 \rrbracket$ code (red squares), and its corresponding physical $X$ ($Z$) operator is then mapped to the logical operator $\overline{X}_1$ ($\overline{Z}_1$) of the $\llbracket 4,2,2 \rrbracket$ code. Note that the $\llbracket 4,2,2 \rrbracket$ codes in adjacent rows or columns are rotated by $90^\circ$ to maintain the orientations of logical operators. Therefore, this mapping transforms the $X$ ($Z$) checks of the surface code into the blue (green) octagonal $X$ ($Z$) checks of the color code, as shown in Fig.~\ref{fig:mapping_sc_cc}(d). 

Furthermore, we duplicate the surface code to support the full mapping, where the second green copy differs from the first blue one by a transversal Hadamard rotation. Under the duplication, with the Hadamard rotation, we have the following relations between pairs of logical operators of the $\llbracket 4,2,2 \rrbracket$ codes, namely $\overline{X}_1\equiv \overline{Z}_2$ and $\overline{Z}_1\equiv \overline{X}_2$. This ensures the syndrome correspondence in Fig.~\ref{fig:mapping_sc_cc}(b,c) such that all depolarizing errors on the surface codes now have a one-to-one mapping with a two-qubit bit-flip error on the color code.
The resulting mapping from the blue surface code to the color code is as follows:
\begin{itemize}
    \item A $Z$ error that violates two adjacent $X$ checks on the blue surface code (Fig.~\ref{fig:mapping_sc_cc}(c)(i)), is mapped to an $\overline{X}_2$ error that violates two adjacent blue checks on the color code (Fig.~\ref{fig:mapping_sc_cc}(b)(i)).
    \item An $X$ error that violates two adjacent $Z$ checks on the blue surface code (Fig.~\ref{fig:mapping_sc_cc}(c)(ii)), is mapped to an $\overline{X}_1$ error that violates two adjacent green checks on the color code, (Fig.~\ref{fig:mapping_sc_cc}(b)(ii)).
    \item A $Y$ error that violates two adjacent $X$ and $Z$ checks on the blue surface code (Fig.~\ref{fig:mapping_sc_cc}(c)(iii)), is mapped to an $\overline{X}_1\overline{X}_2$ error that violates two adjacent blue and green checks on the color code (Fig.~\ref{fig:mapping_sc_cc}(b)(iii)).

\end{itemize}

\subsection{\label{subsec:decoding}Decoding problem}
In quantum error correction, we measure all stabilizer generators to extract the syndrome information for detecting errors. The decoding problem then becomes that of inferring a probable error configuration consistent with the observed syndromes. Before diving into the decoding problem, we first define the quantum memory noise models considered in this work, where errors are typically assumed to be independent and identically distributed (i.i.d.):
\begin{itemize}
    \item \textbf{Code capacity noise model}: Errors may occur on physical data qubits of the code at each stabilizer measurement step, while the measurements are assumed to be perfect.
    \item \textbf{Phenomenological noise model}: Errors may affect both data qubits and ancilla qubits at each stabilizer measurement round. Measurement errors on ancilla qubits flip their syndrome outcomes.

\end{itemize}

For simplicity, we focus on CSS codes~\cite{calderbank1996good, steane1996multiple}, whose stabilizers consist exclusively of either $X$- or $Z$-type operators, allowing for the independent and analogous decoding of $X$-type and $Z$-type Pauli errors. In what follows, we describe the decoding procedure for $X$/$Z$-errors.
The decoding problem aims to infer the most probable error $e'$ given the syndrome $s$. An optimal decoder employs maximum likelihood estimation, solving $\max \sum_{e\in f(e')} P(e \mid s)$, where $f(e') = e' + \mathcal{S}$ denotes the coset of errors equivalent to $e'$ up to the addition of any stabilizer. However, optimal decoding is generally \#P-complete
~\cite{iyer2015hardness}, making it computationally intractable. Thus, practical quantum computing requires alternative decoders that trade optimality for efficiency. One widely used decoder for topological codes, such as the surface and color codes, is the minimum-weight perfect matching (MWPM) algorithm. The MWPM decoder identifies an approximate maximum likelihood error $e'$ that maximizes $P(e' \mid s)$ by matching violated checks on a matching graph $G = (V, E)$ constructed from the parity-check matrix $H$. Here, $V$ represents the set of checks, and $E$ represents the data qubits. Each edge is assigned a weight as $\log[(1 - q)/q]$, where $q$ is the error probability associated with that edge. 

The above description assumes the code capacity noise model. Under the phenomenological noise model, measurement errors $\epsilon$ on ancilla qubits modify the syndrome as $s = H e^T + \epsilon^T$. In practice, multiple rounds ($r$) of stabilizer measurements are performed to identify measurement errors.
The syndrome in round $i$ is given by $s_i=He_i^T+\epsilon_i^T \text{ for }i=0,1,...,r \text{, with }\epsilon_r=0$. This leads to the consolidated equation $\mathbf{G} \mathbf{e}^T = \mathbf{s}^T$, where $\textbf{G}=(I_m\otimes H|R \otimes I_n)$, $\textbf{e}=(e_1,e_2,...,e_r|\epsilon_1,\epsilon_2,..,\epsilon_{r})$, and $\textbf{s}=(0+s_1,s_1+s_2,...,s_{r-1}+s_r)$. Here, $\mathbf{G}$ can be interpreted as the parity-check matrix of a hypergraph product code combining the stabilizer code with a repetition code. Instead of using the raw syndrome, decoding is performed on the syndrome differences between consecutive rounds. The matching graph in this case becomes a space-time graph, capturing both spatial and temporal error correlations.

\section{\label{sec:correlated decoder}The Correlated Decoder} 

\begin{figure*}[ht]
    \centering
    \includegraphics[width=1\linewidth]{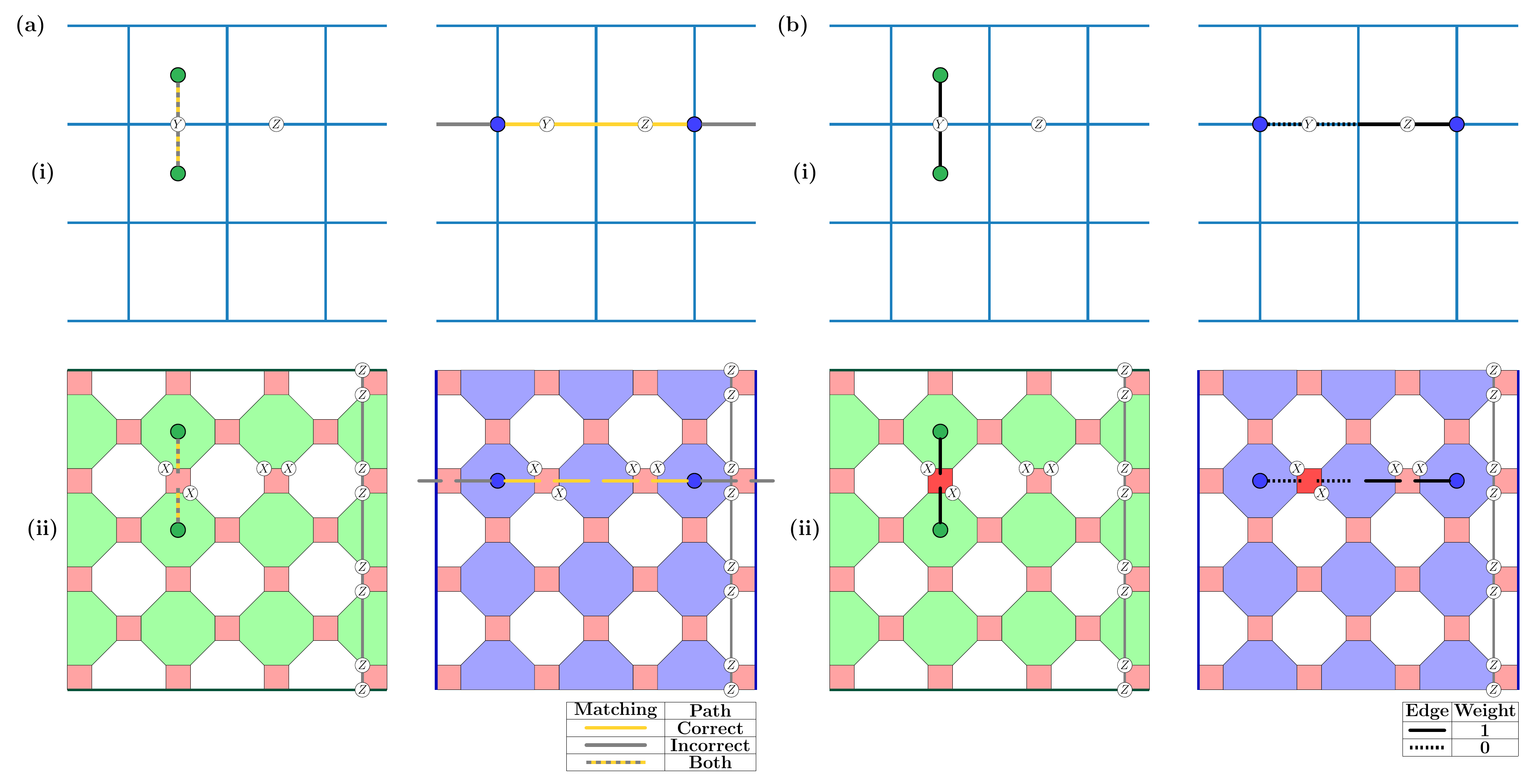} 
    \caption{The restricted and correlated decoders for the color code can be derived from their surface code counterparts via the color-surface mapping. (a) The independent $X/Z$ decoder (i) and the restricted decoder (ii) exhibit an analogous limitation: while decoding the degenerate weight-$O(d/2)$ error patterns (containing $Y$ or diagonal errors), they overestimate the matching weight of the correct path, results in a success probability of only 1/2. (b) This issue is resolved by the correlated decoders for both codes (i, ii) by setting the weight of edges identified in the first matching to zero during the second matching. This ensures that the predicted error weight is consistent with the total matching weight, thus achieving more accurate decoding. 
   }
    
    \label{fig:correlated_decoding}
\end{figure*}

The conventional restricted decoder for the 4.8.8 color code functions by independently decoding on two of the three restricted lattices, $\mathcal{R}_\textbf{b}$ and $\mathcal{R}_\textbf{g}$. Here, $\mathcal{R}_\mathbf{u}$ denotes the restricted lattice obtained by removing all color-$\mathbf{u}$ checks.
A limitation of this approach is that it cannot correct the certain weight-$O(d/2)$ error configurations shown in Fig.~\ref{fig:correlated_decoding}(a)(ii). While decoding on $\mathcal{R}_\textbf{g}$ lattice, the restricted decoder matches syndrome defects along one of the correct and the incorrect paths with equal weights, making a successful probability of 50\%. However, this issue can be resolved by rectifying the overestimated weights, which is the key improvement introduced by the correlated decoder.

Let's re-examine the mapping in Sec.~\ref{sec:Mapping}. This mapping transforms the independent $X$/$Z$ decoding for the surface code against depolarizing error into the independent $\mathcal{R}_\mathbf{b}$/$\mathcal{R}_\mathbf{g}$ restricted lattice decoding for the color code against two-qubit bit-flip error (i.e., the restricted decoder), illustrated in Fig.~\ref{fig:correlated_decoding}(a). Consequently, the drawback of the restricted decoder mirrors that of the independent $X$/$Z$ decoding for the surface code. For decoding depolarizing errors on the surface code, the independent $X$/$Z$ decoder cannot resolve the degeneracy between a combined $X$-$Z$ error and a $Y$ error on a qubit, causing a overestimated total weight-2 matching over $X$ and $Z$ decoding for a weight-1 $Y$ error. Consequently, the $X$/$Z$ independent decoder correctly decode the error shown in Fig.\ref{fig:correlated_decoding}(a)(i) with a probability of 1/2. Analogously from the mapping, the restricted decoder may overestimate the weight of the correct matching path when presented with diagonal error configurations as shown in Fig.\ref{fig:correlated_decoding}(a)(ii). This is because it cannot resolve the degeneracy between a combined vertical-horizontal weight-2 $X$ error and a diagonal weight-2 $X$ error on a red square. The correlated MWPM decoder for surface code addresses this issue by setting the weight of edges identified in the first matching to zero during the second matching, as shown in Fig.~\ref{fig:correlated_decoding}(b)(i). This adjustment ensures that the total weight of the predicted error aligns with the overall matching weight, which is crucial for MWPM-based decoders to achieve optimal performance. 

\begin{figure}[ht]
    \centering
    \includegraphics[width=1\linewidth]{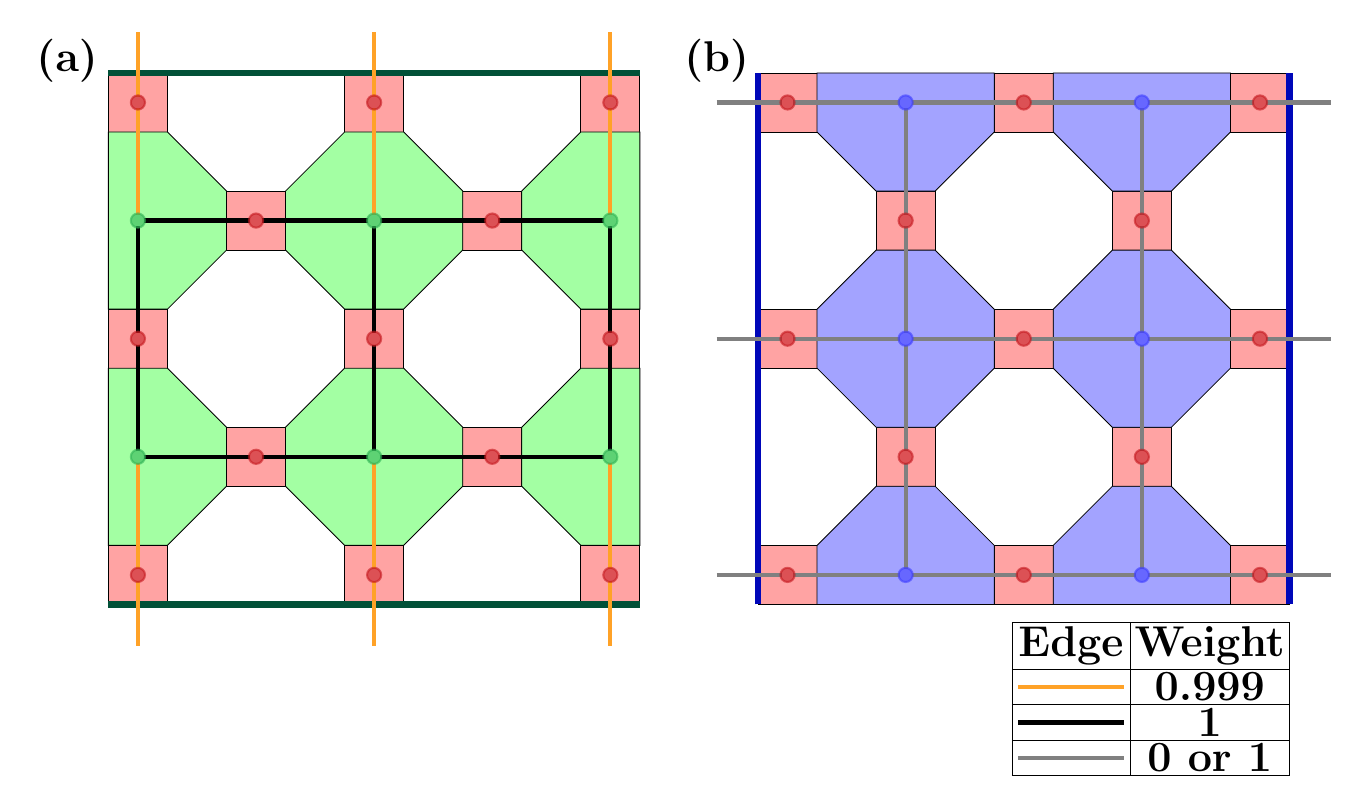}
    \caption{Matching graph of the correlated decoder. Dangling edges are connected to a virtual boundary check, defined as the product of all the checks. (a) In $\mathcal{R}_\textbf{b}$ lattice, edges adjacent to the top and bottom red checks are assigned a weight of 0.999 to handle boundary errors, while others have weight 1. (b) In $\mathcal{R}_\textbf{g}$ lattice, edge weights are set to 0 or 1 depends on the decoding results from the $\mathcal{R}_\textbf{g}$ lattice.}
    
    \label{fig:decoding_graph}
\end{figure}

By applying this mapping to the correlated decoder for surface code, we can derive the correlated decoder for the color code to address analogous issues. The corresponding matching graph is illustrated in Fig.~\ref{fig:decoding_graph}. The core principle involves setting the weight of edges identified in the $\mathcal{R}_\textbf{b}$ matching to zero during the $\mathcal{R}_\textbf{g}$ matching. We further make a modification to handle boundary errors: the weights of edges adjacent to the top and bottom red checks in the $\mathcal{R}_\mathbf{b}$ lattice are set to 0.999, which is detailed in Section~\ref{sec:performance analysis}. This correlated decoder successfully corrects the aforementioned weight-$O(d/2)$ error shown in Fig.~\ref{fig:correlated_decoding}(b) by reducing the total weight of the correct matching to 4, equal to the actual error weight. For decoding the logical qubit defined by the vertical logical $Z$-operator (the two restricted lattices exchanges for the other qubit), the decoding procedure of the correlated decoder is as follows:

\begin{enumerate}
    \item Decode on the $\mathcal{R}_\textbf{b}$ lattice and mark the red checks (the deeper red squares in Fig.~\ref{fig:correlated_decoding}) that are completely traversed by matching edges.
    \item Set the weights of the edges in the $\mathcal{R}_\textbf{g}$ lattice (dashed lines in Fig.~\ref{fig:correlated_decoding}) connected to the marked red checks to 0, then proceed to decode on the $\mathcal{R}_\textbf{g}$ lattice.
\end{enumerate}

\begin{figure}[ht]
    \centering
    \includegraphics[width=1\linewidth]{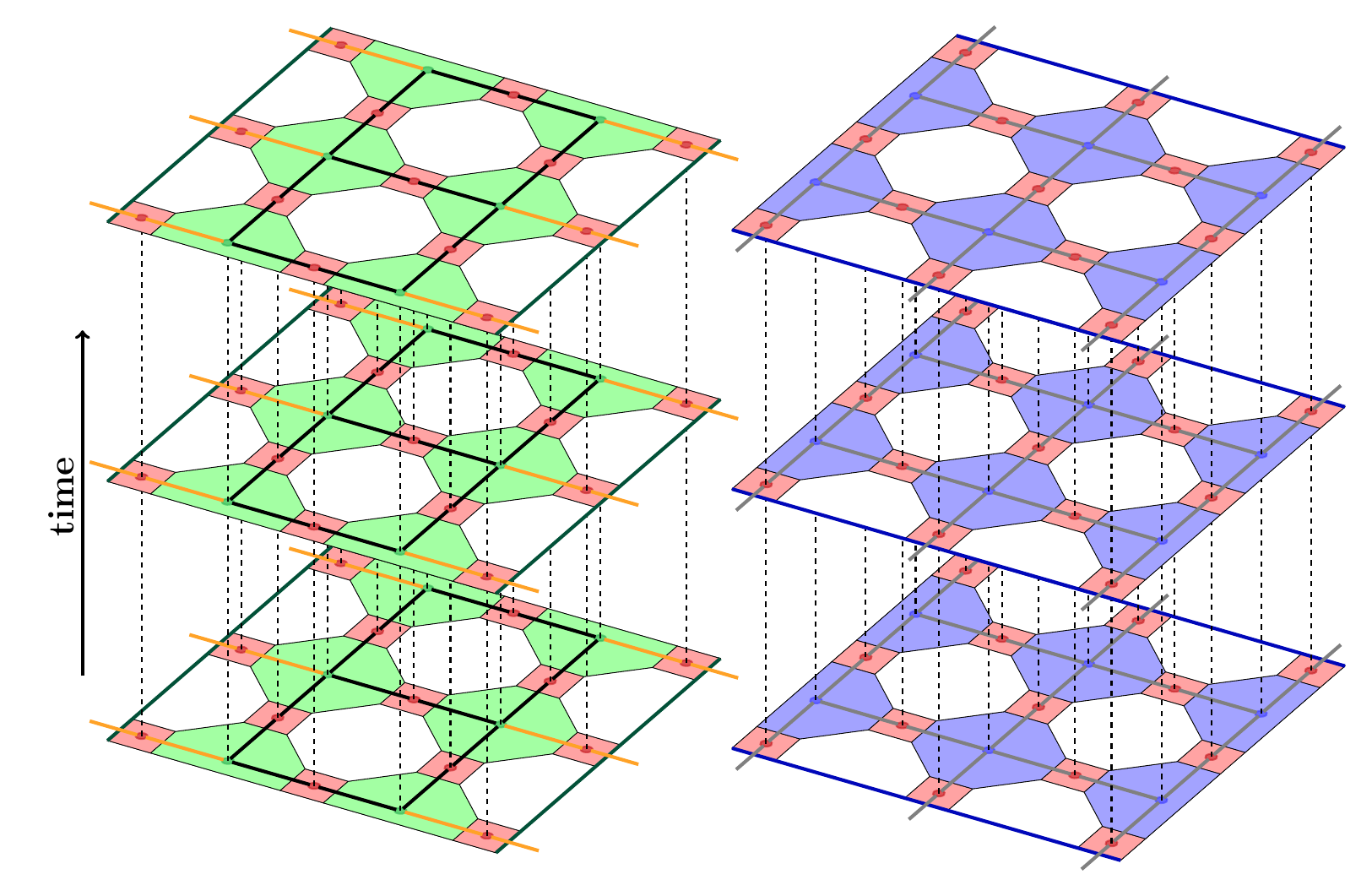}
    \caption{ The space-time matching graph for $r=3$ rounds under the phenomenological noise model. Time progresses upward from the bottom, with each horizontal plane corresponding to a round of stabilizer measurements. The vertical dashed edges represent space-time locations where measurement errors may occur.
    }
    \label{fig:3d_decoding}
\end{figure}

The correlated decoder incurs no additional computational overhead compared to the restricted decoder, but it sacrifices the parallelism of the two decoding stages.
For the phenomenological noise model, the correlated decoding procedure is adapted by applying the same two-step process on the space-time matching graphs, as illustrated in Fig.~\ref{fig:3d_decoding}.

\section{\label{sec:performance analysis}Performance Analysis at Low Error Rates}
In the regime of asymptotically low physical error rates, the logical error rate is mainly determined by the number of minimum-weight failure mechanisms 
\vspace{-1ex}
\begin{equation}
P_{\text{fail}} 
\sim  N_{\text{fail}}\, p^{m} (1-p)^{n-m} \sim  N_{\text{fail}}\, p^{m}    
\end{equation}
where $N_{\text{fail}}$ is the number of minimum-weight ($m$) error configurations  leading to a logical failure. The low error-rate performance of the restricted and unified decoders has been thoroughly analyzed in Ref.~\cite{benhemou2025minimising} where the number of minimum-weight failure configurations for decoding the color code is given by
\vspace{-1ex}
\begin{equation}
N_{\textrm{fail}}^{\textrm{res}}  = \frac{M}{2}\sum_{k=0}^{\left \lfloor \frac{M}{2}\right \rfloor}\begin{pmatrix}
 M\\k
\end{pmatrix}\begin{pmatrix}
 M-k\\k
\end{pmatrix} 4^{M-k}
\label{eq:N_fail_color_restricted} 
\end{equation}
\vspace{-2ex}
\begin{equation} 
N_{\textrm{fail}}^{\textrm{uni}} = \frac{M}{2}\sum_{k=0}^{\left \lfloor \frac{M}{2}\right \rfloor}\begin{pmatrix}
 M\\k
\end{pmatrix}\begin{pmatrix}
 M-k\\k
\end{pmatrix} 2^{k}4^{M-2k}
\label{eq:N_fail_color_unified}
\end{equation} 
where $M$ denotes the number of rows supporting $d/2$ red square plaquettes in the lattice, $k$ is the number of red plaquettes raising a weight-2 error, and $M-2k$ is the number of weight-1 errors. The failure configurations of the restricted decoder consist of all weight-$d/2$ errors aligned along a single row, with the prefactor $1/2$ accounting for the probability of the matching decoder correctly guessing the error. In contrast, the unified decoder excludes any error patterns containing a two-qubit diagonal error from these failure configurations.

\begin{figure}[b]
    \centering
    \includegraphics[width=\linewidth]{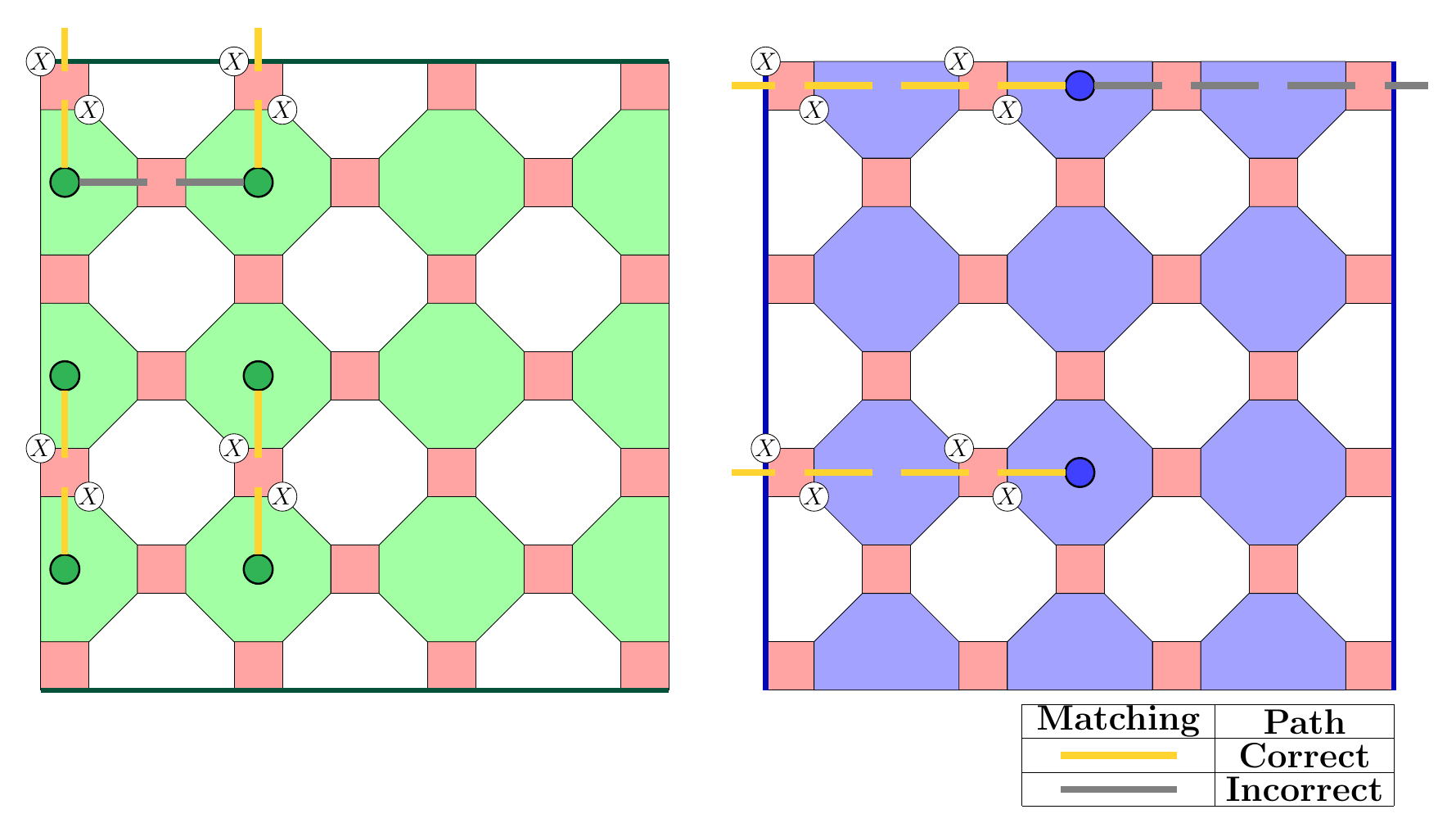}
    \caption{Comparative difficulty of correcting errors on the boundary versus the interior of the lattice. Errors in the bulk (middle) are correctable by the correlated decoder, whereas identical error configurations on the boundary have a probability 1/2 of logical failure.}
    \label{fig:performance_analysis}
\end{figure}

\begin{figure}[th]
    \centering
    \includegraphics[width=\linewidth]{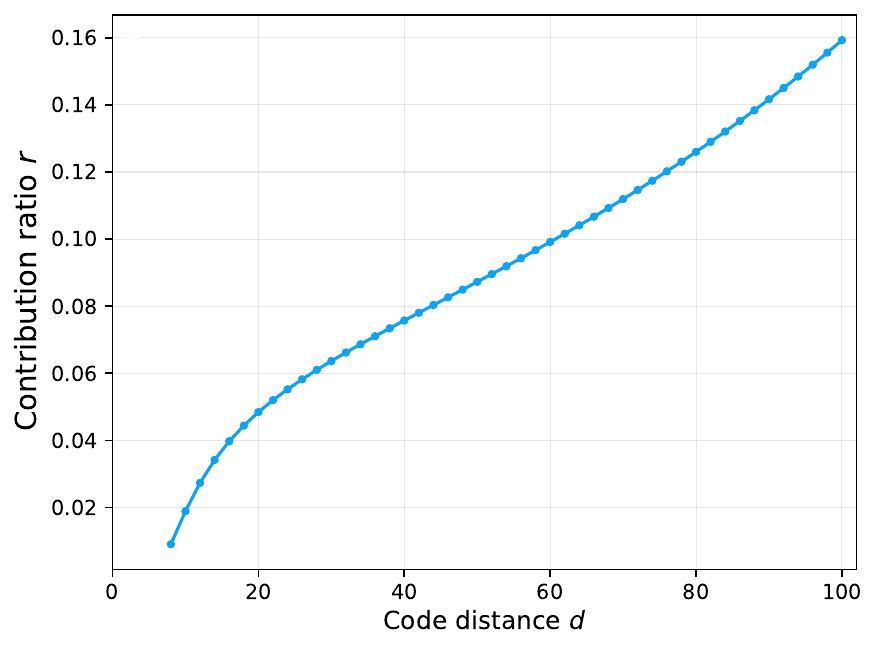}
    \caption{Relative contribution $r=N_{\textrm{b}}/N_{\textrm{fail}}^{\textrm{uni}}$ of  additional boundary failure mechanisms to the total failure configurations of the unified decoder, evaluated over code distance $d$ ranging from 8 to 100.}
    \label{fig:contribution}
\end{figure}

The correlated decoder achieves the performance identical to the unified decoder for weight-$d/2$ errors located on the middle rows of the lattice. However, on the boundary rows (the first and last rows), a small fraction of the otherwise correctable error configurations involving diagonal errors, become uncorrectable due to the reduced check connectivity, as illustrated in Fig.~\ref{fig:performance_analysis}.
To mitigate this, we slightly reduce the weights of the edges adjacent to the checks incident to the green boundary in $\mathcal{R}_\textbf{g}$ lattice, thereby introducing a preference for boundary corrections in the correlated decoder. While this adjustment is heuristic, numerical evidence confirms that it effectively eliminates dominant boundary failure mechanisms without introducing new ones, leading to an improvement in the thresholds. We provide a comprehensive set of supporting evidence in Appendix.~\ref{sec:appendix_A}, including the error configurations examples that can be corrected through this adjustment (Fig.~\ref{fig:solve_boundary_error}) and a threshold comparison with and without the adjustment (Fig.~\ref{fig:thresholds_cmp}). After applying the adjustment, the remaining boundary failures are characterized by adjacent diagonal error pairs as shown in Fig.~\ref{fig:remaining_2D_error} of Appendix.~\ref{sec:appendix_A}, and quantified by $N_{\textrm{b}}$:

\vspace{-1ex}
\begin{equation}
\begin{split}
N_{\textrm{b}}  = & \frac{1}{2}\times 2 \sum_{i=1}^{\left\lfloor \frac{M}{4} \right \rfloor}
\sum_{j=0}^{\left \lfloor \frac{M}{2}-2i \right \rfloor} 
\binom{M-i}{i} \binom{M-2i}{j} \binom{M-2i-j}{M-4i-2j}\\
& \times 4^i \times 2^j \times 4^{M-4i-2j} \\
= & \sum_{i=1}^{\left\lfloor \frac{M}{4} \right \rfloor}
\sum_{j=0}^{\left \lfloor \frac{M}{2}-2i \right \rfloor} 
\binom{M-i}{i} \binom{M-2i}{j} \binom{M-2i-j}{M-4i-2j} \\
 &\times 4^{M-3i-\frac{3}{2}j}
\end{split}
\label{eq:N_extra_color_correlated} 
\end{equation}
where $i$ is the number of adjacent diagonal error pairs, and $j$ is the number of a weight-2 error located on the top/bottom edge of a square plaquette. The total number of minimum-weight failure configurations for the correlated decoder is thus:
\begin{equation}
N_{\textrm{fail}}^{\textrm{cor}} = N_{\textrm{fail}}^{\textrm{uni}}+N_{\textrm{b}}
\label{eq:N_fail_color_correlated} 
\end{equation}
The relative contribution $r=N_{\textrm{b}}/N_{\textrm{fail}}^{\textrm{uni}}$ of additional boundary failure mechanisms is plotted as a function of code distance $d$ in Fig.~\ref{fig:contribution}. Accordingly, the logical error rate of the correlated decoder exceeds that of the unified decoder by a factor of $1+r$. Given that $r$ remains small for practical code distances, the performance of the correlated decoder is closely approximates that of the unified decoder in the low error-rate regime.

Since the computational complexity of MWPM scales as $O(|V|^2\times |E|)$ for the matching graph $(V,E)$~\cite{edmonds1965paths, higgott2022pymatching}, making graph size critical for decoding time. The correlated decoder performs MWPM on two restricted lattices, while the unified decoder perform MWPM on a unified lattice connected by three restricted lattices. Consequently, while they both exhibit the same asymptotic complexity, the unified decoder incurs approximately 13.5 times higher computational overhead than the correlated decoder, which makes the correlated decoder more efficient than the unified decoder.

\section{\label{sec:results}Numerical results}
\begin{figure}[th]
    \centering
    \includegraphics[width=\linewidth]{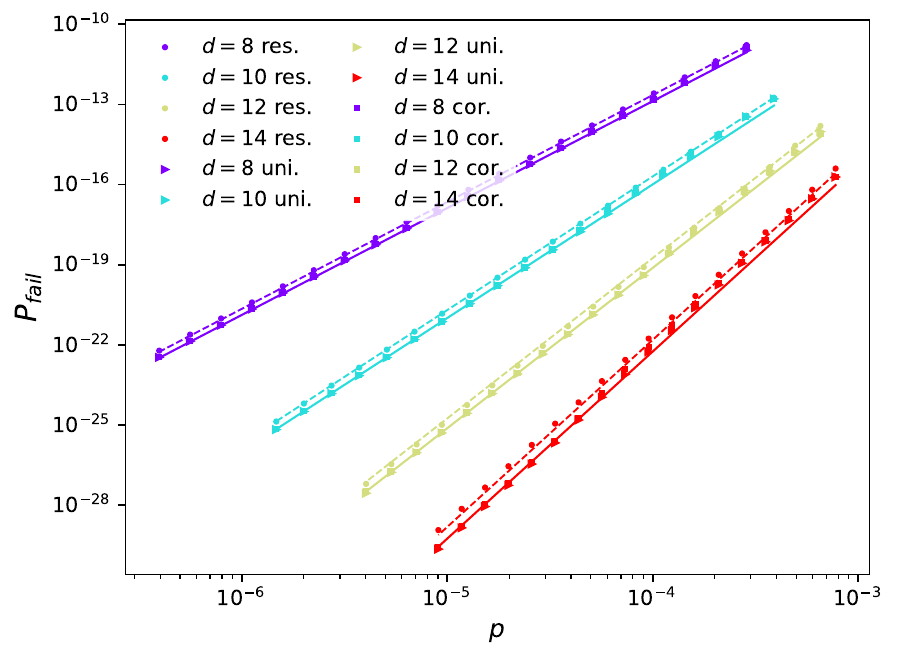}
    \caption{Logical failure rates at low physical error rates (logarithmic scale) 
    for the restricted, unified and correlated decoders on the color code under bit-flip noise. Dashed and solid lines represent the analytical predictions for the restricted and unified decoders, respectively, based on path-counting methods~\cite{benhemou2025minimising}.
    }
    \label{fig:splitting}
\end{figure}

To numerically evaluate the logical failure rates at low physical error rates, we use the rare-event splitting method introduced in Ref.~\cite{bravyi2013simulation}, follow the implementation of Ref.~\cite{benhemou2025minimising}. Fig.~\ref{fig:splitting} presents a comparison of the logical failure rates for the three decoders. The numerical results demonstrate good agreement with our analytical estimates. 

\begin{figure*}[htb]
    \centering
    \begin{subfigure}{0.48\linewidth}
        \centering
        \includegraphics[width=\linewidth]{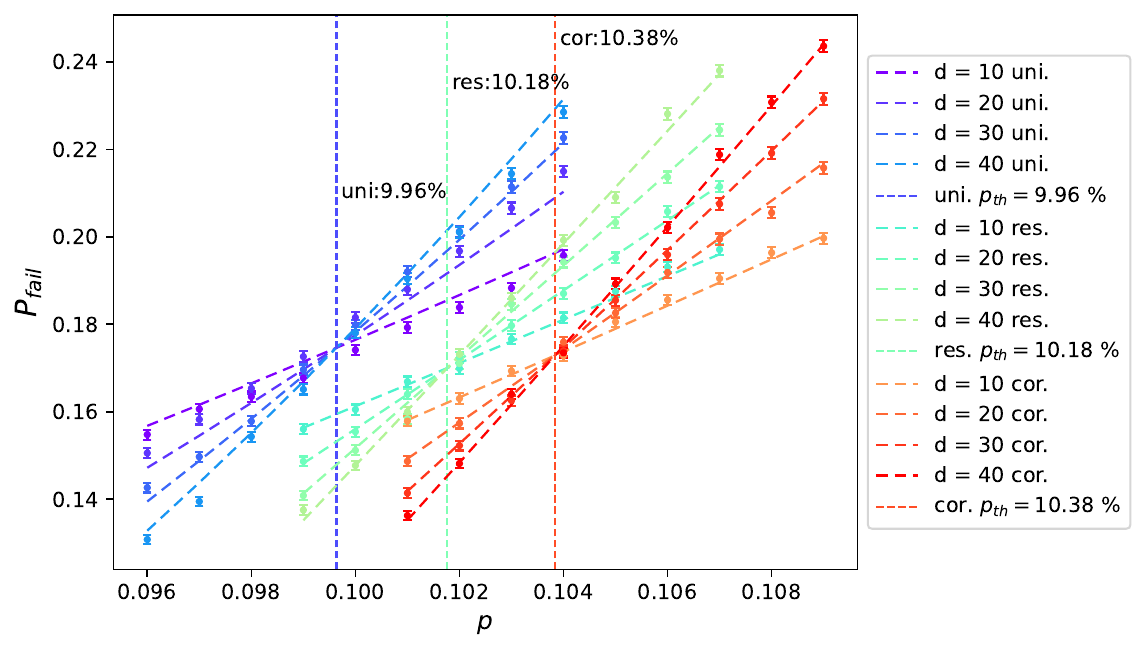}
        \caption{Threshold plot for decoding the color code against bit-flip error under code capacity noise model.}
        \label{fig:color_cap}
    \end{subfigure}
    \hfill
    \begin{subfigure}{0.48\linewidth}
        \centering
        \includegraphics[width=\linewidth]{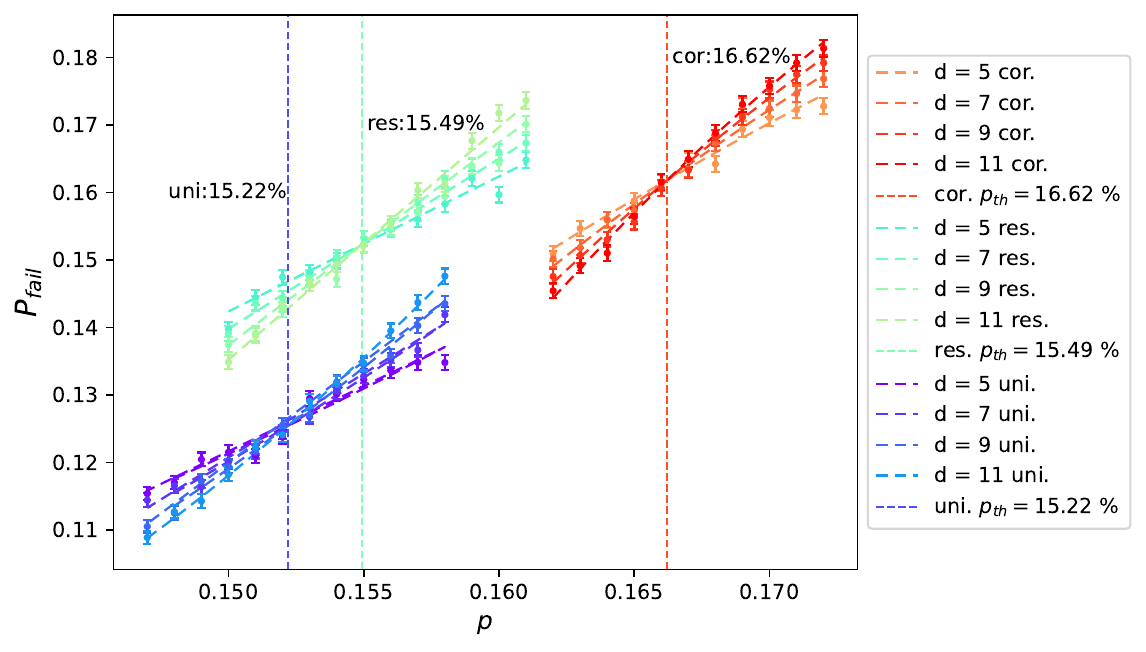}
        \caption{Threshold plot for decoding the surface code against depolarizing error under code capacity noise model.}
        \label{fig:toric_cap}
    \end{subfigure}
    \vspace{0.5cm}
        \noindent
    \begin{subfigure}{0.48\linewidth}
        \centering
        \includegraphics[width=\linewidth]{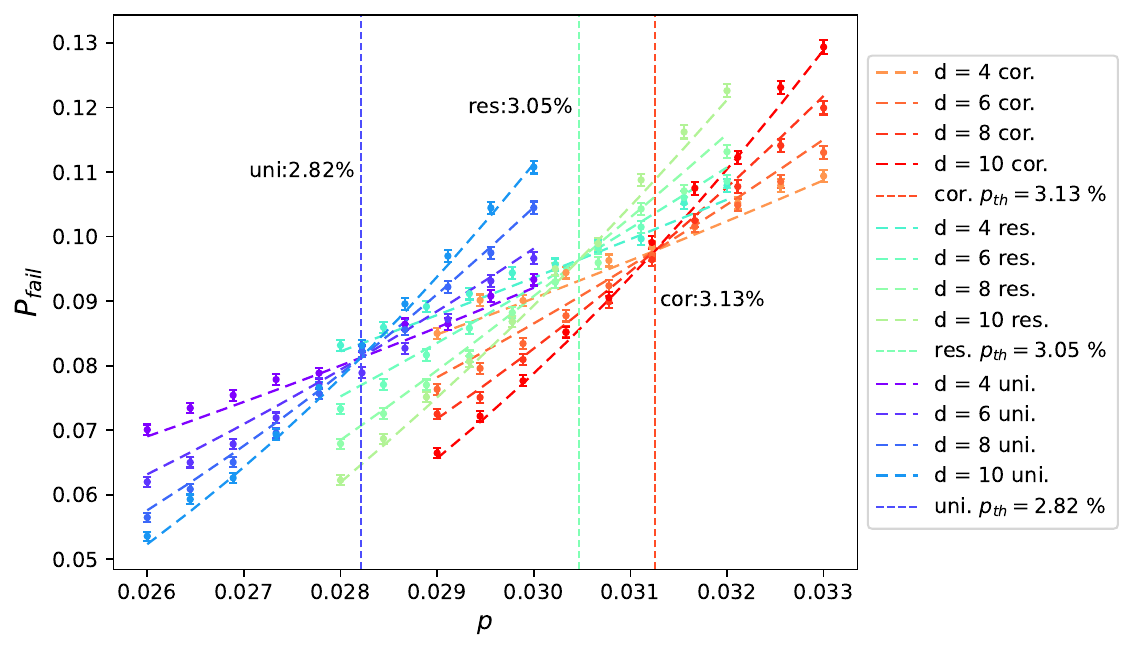}
        \caption{Threshold plot for decoding the color code against bit-flip error under phenomenological noise model.}
        \label{fig:color_phenom}
    \end{subfigure}
    \hfill
    \begin{subfigure}{0.48\linewidth}
        \centering
        \includegraphics[width=\linewidth]{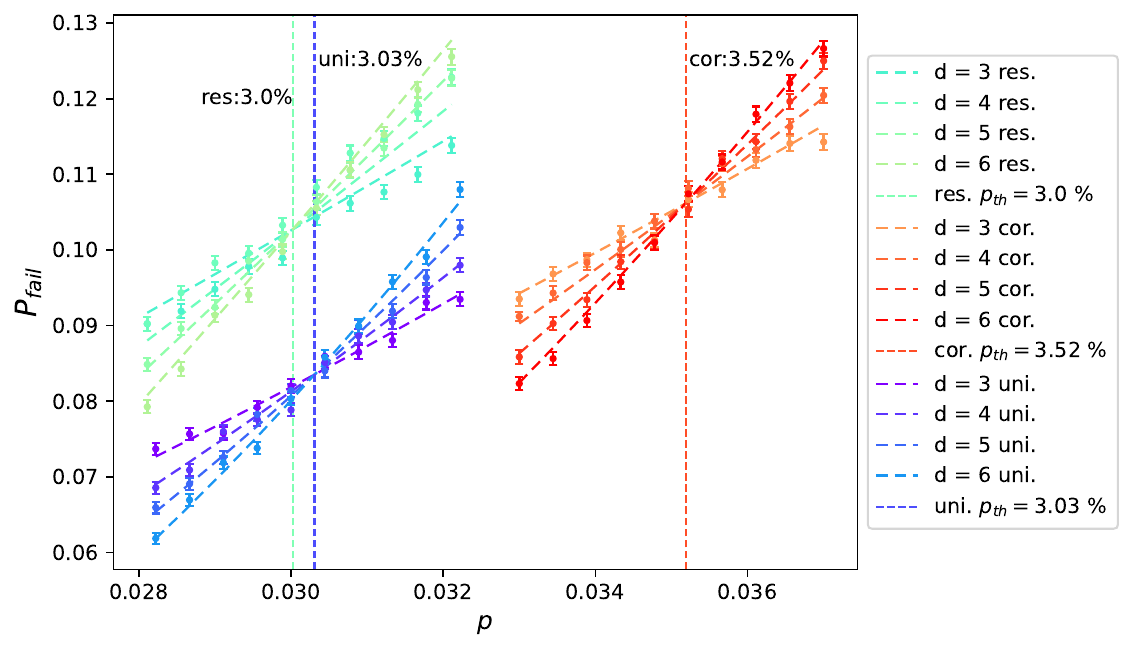}
        \caption{Threshold plot for decoding the surface code against depolarizing error under phenomenological noise model. }
        \label{fig:toric_phenom}
    \end{subfigure}
    
    \caption{Thresholds comparison of the restricted, unified and correlated decoders. The logical failure rate $P_{\text{fail}}$ is plotted as a function of physical error rate $p$ for different distances $d$. The threshold is indicated by the intersection of the curves. The error bars show the standard deviation of the mean logical failure rate, with each data point obtained from $10^5$ Monte Carlo samples. The data collected is fit to the Taylor function $f=Ax^2+Bx+C$, where $x$ is the rescaled error rate $x=(p-p_{th})^{1/v}$. The number of stabilizer measurement rounds $r$ is set to the distance $d$ under the phenomenological noise model.
    }
    \label{fig:thresholds}
\end{figure*}
We estimate the thresholds of our proposed correlated decoder using Monte Carlo simulations and make a comparison with the restricted and unified decoders. The threshold is calculated by fitting the logical error rate data to a Taylor expansion truncated to the quadratic term $f=Ax^2+Bx+C$, where $x=(p-p_{th})^{1/v}$ is the rescaled error rate. This method is detailed in Ref.~\cite{wang2003confinement}. 

As shown in Fig.~\ref{fig:thresholds}, the results demonstrate that the correlated decoder achieves the thresholds of 10.38\% and 16.62\% for decoding the color and surface code under code capacity noise model, respectively. Furthermore, we find a threshold of 3.13\% and 3.52\% for decoding the color and surface code under phenomenological noise model. The thresholds demonstrate the advantage of the correlated decoder over the restricted and unified decoders.

\section{\label{sec:discussion}Discussion}

In this work, we introduce a correlated decoder for the 4.8.8 color code, which enhances the restricted decoder by leveraging correlations between the restricted lattices. Through the color-surface code mapping, the correlation we leverage is precisely the $X$/$Z$ correlation required for decoding surface codes under depolarizing noise, thereby instantiating the general correlated CSS decoding framework established in Ref.~\cite{delfosse2014decoding_cc}. We evaluate the decoder's performance at low physical error rates analytically and numerically. Our results demonstrate a significant improvement in the threshold over the restricted decoder, and an exponential suppression of the logical error rates for the color code under bit-flip noise.

The performance advantage of our decoder stems from a fundamental principle in MWPM-based decoding. Achieving optimal performance with MWPM-based decoders requires consistency between the weight of the decoded errors and matching edges. This consistency is naturally maintained in MWPM decoders for surface codes due to the direct correspondence between edges and errors. In color codes, however, the higher connectivity of qubits and the increased degeneracy of error representations often break this consistency. For instance, in the projection decoder for the 6.6.6 lattice, the final error is determined by filling the cycle enclosed by all the matching edges on three restricted lattices, which can yield an error whose weight (after simplification by stabilizers) does not reliably reflect the matching weight. This explains the previously reported limitation in decoding weight-$O(d/3)$ errors~\cite{sahay2022decoder}. Our correlated decoder substantially mitigates this "inconsistent weight" problem in the 4.8.8 color code compared to the restricted decoder. 

Although we explored generalizing the correlated decoder to the 6.6.6 lattice, we found that it could not address the fundamental limitation of correcting the weight-$O(d/3)$ errors inherent to its projection decoder.
Nevertheless, the core idea of leveraging the correlations of restricted lattices to resolve weight inconsistency is broadly valuable, which is supported by prior work~\cite{lee2025color}. Extending this principle to develop correlated decoders for other code families as well as for circuit-level noise models, constitutes a promising direction for future research.

\begin{acknowledgments} 
The authors thank Benjamin J Brown and Ying Li for valuable feedback and Weilei Zeng for useful discussions. 
This work has been supported by the National Key R\&D Program of China (Grant No. 2024YFB4504001), the National Natural Science Foundation of China (Grant No. 62302395), and the Aid Program for Science and Technology Innovative Research Team in Higher Educational Institutions of Hunan Province.
\end{acknowledgments}

\bibliography{ref.bib}

\appendix
\section{\label{sec:appendix_A}Numerical analysis of boundary errors}
The correlated decoder successfully corrects the degenerate weight-$d/2$ error patterns involving diagonal errors within the middle rows of the lattice. However, a subset of such errors located on the boundary rows (i.e., the first and last rows) becomes more difficult to correct due to the reduced check connectivity. Here we provide a detailed analysis of these boundary errors. 

\begin{figure*}[htb]
    \centering
    \begin{minipage}{0.48\linewidth}
        \centering
        \includegraphics[width=\linewidth]{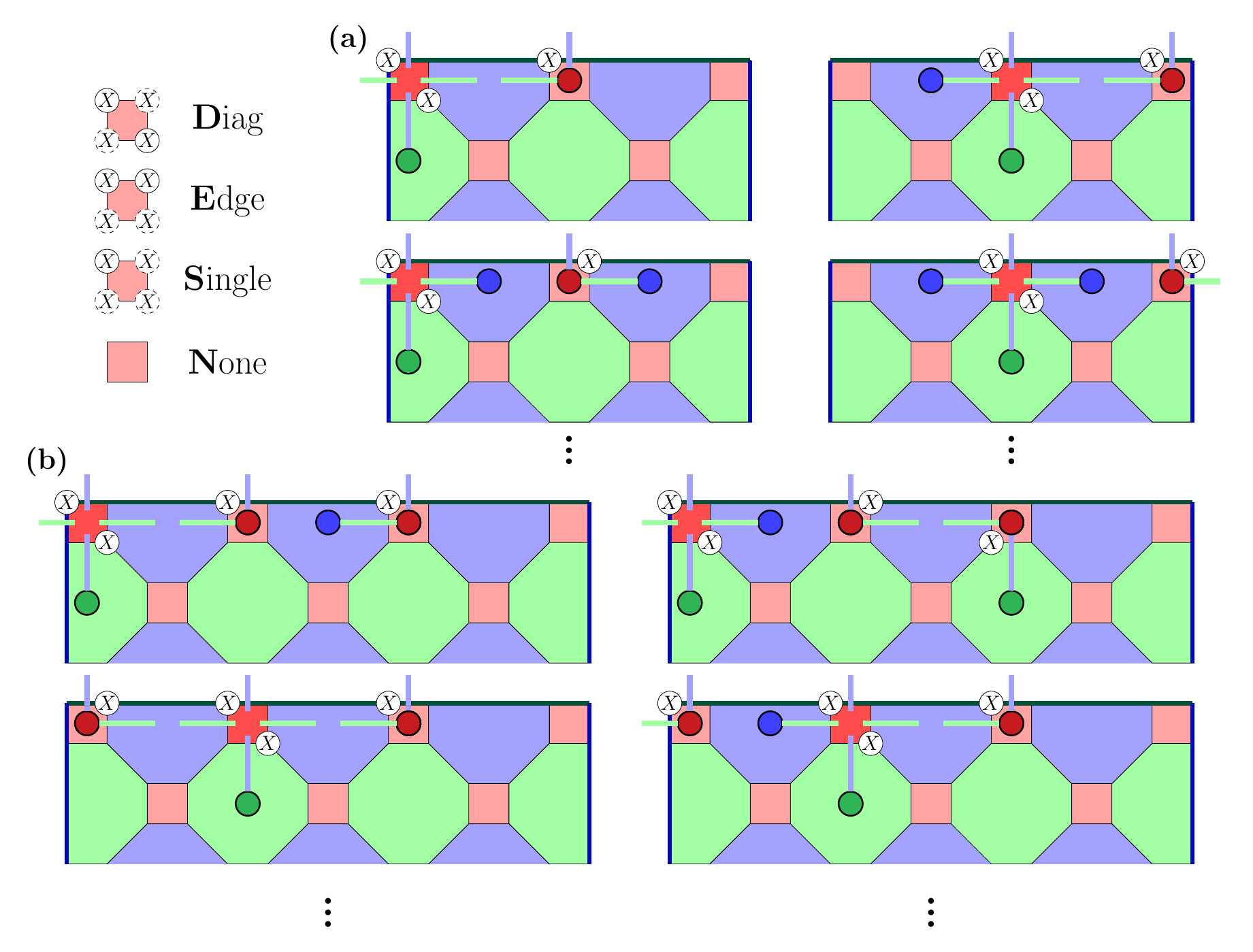}
        \caption{Examples of the error configurations that become correctable after modifying $w_b$ to 0.999. Blue and green matchings indicate correct matchings in the $\mathcal{R}_\mathbf{g}$ and $\mathcal{R}_\mathbf{b}$ lattices, respectively.}
        \label{fig:solve_boundary_error}
    \end{minipage}
    \hfill
    \begin{minipage}{0.48\linewidth}
        \centering
        \includegraphics[width=\linewidth]{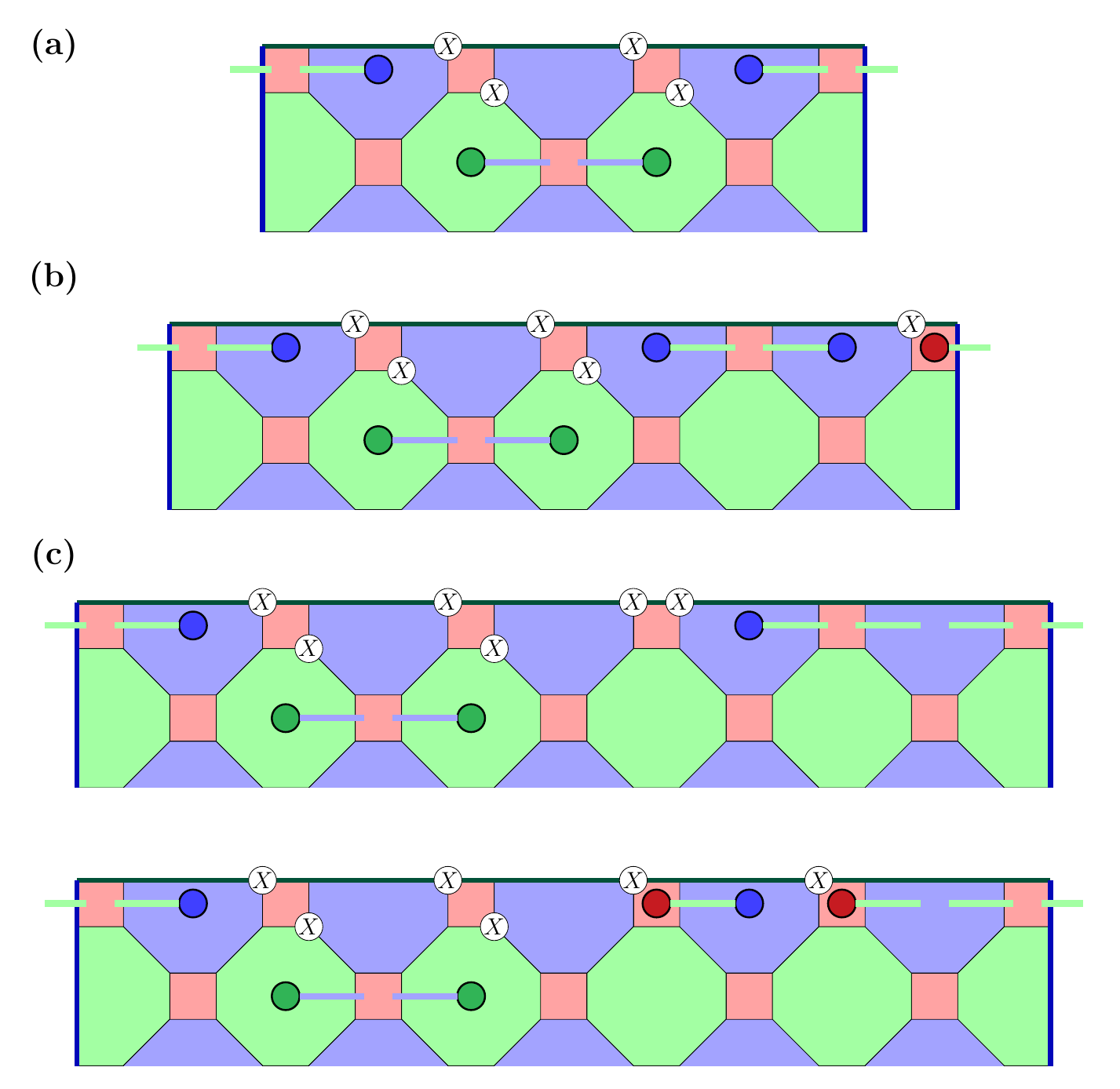}
        \caption{Examples of remaining failure configurations consisting of adjacent diagonal error pairs patterns for $d=8,10,12$. Blue and green matchings indicate failed matchings in the $\mathcal{R}_\mathbf{g}$ and $\mathcal{R}_\mathbf{b}$ lattices, respectively, occurring with the probability of $1/2$.}
        \label{fig:remaining_2D_error}
    \end{minipage}
\end{figure*}

To support our analysis, we introduce the following notation. Minimum-weight failure configurations along a row of $M=d/2$ red squares consist of four error types per square: Diagonal errors (2 configurations), Edge errors (2 configurations), Single errors (4 configurations), and No error (1 configuration), denoted by their initial letters as illustrated in the top-left of Fig.\ref{fig:solve_boundary_error}. An error pattern is defined as a multiset $\{e_1, e_2, \dots, e_{d/2}\}$ where element order is irrelevant and each $e_i \in \{\text{D, E, S, N}\}$. For example, the pattern $\{\text{D, S, S, N}\}$ corresponds to one diagonal-error square, two single-error squares, and one no-error square in a row. The number of such configurations is given by $\binom{4}{1}\binom{3}{1}\times 2 \times 4 ^2$.

\begin{figure*}[htb]
    \centering
    \begin{subfigure}{0.48\linewidth}
        \centering
        \includegraphics[width=\linewidth]{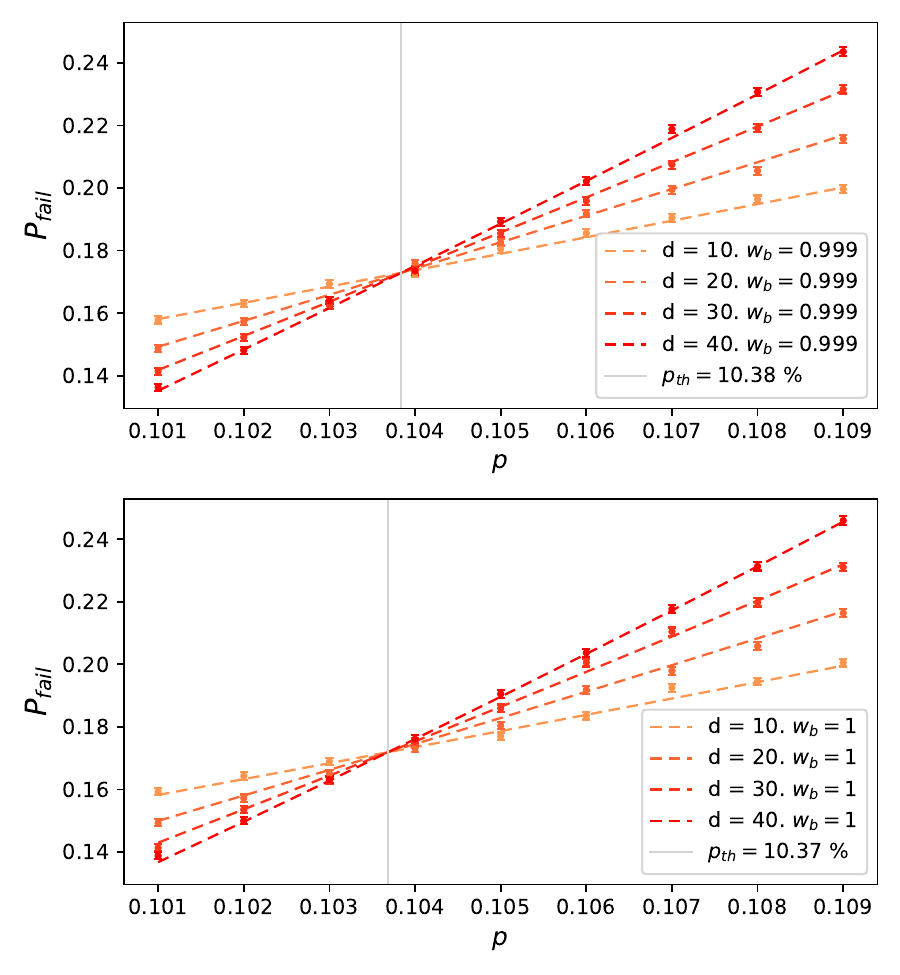}
        \caption{Thresholds of decoding the color code for $w_b=0.999$ (top) and $w_b=1$ (bottom).}
        \label{fig:color_cmp}
    \end{subfigure}
    \hfill
    \begin{subfigure}{0.48\linewidth}
        \centering
        \includegraphics[width=\linewidth]{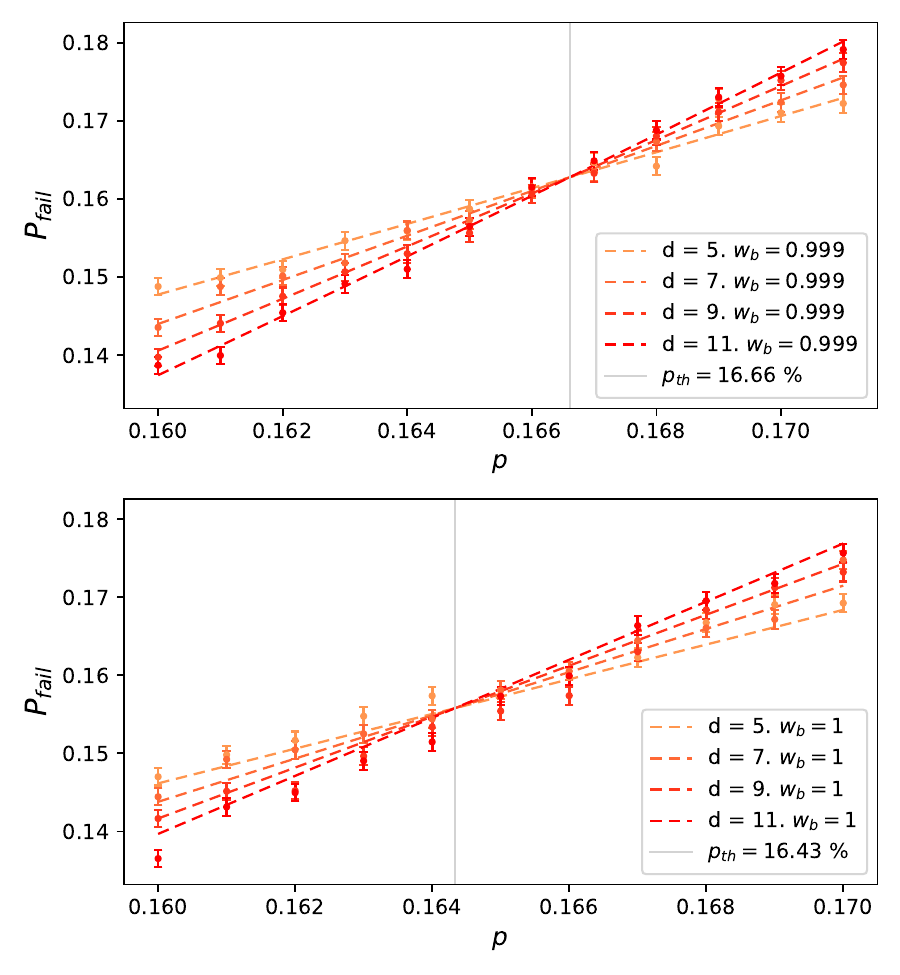}
        \caption{Thresholds of decoding the surface code for $w_b=0.999$ (top) and $w_b=1$ (bottom).}
        \label{fig:toric_cmp}
    \end{subfigure}
    
    \caption{Comparison of decoding thresholds with and without boundary edge weight $w_b$ adjustment.
    }
    \label{fig:thresholds_cmp}
\end{figure*}

By modifying the weight $w_b$ of the edges connected to the top and bottom red checks in the $\mathcal{R}_\mathbf{b}$ lattice from 1 to 0.999, the number of additional boundary failures is dramatically reduced without introducing new minimum-weight failures. As summarized in Tab.~\ref{tab:enumerating_error}, the simulated failure counts $N$ (before adjustment) and $N'$ (after adjustment) show significant reductions for specific error patterns. Fig.~\ref{fig:solve_boundary_error} shows correctable boundary failure examples by this adjustment. Threshold comparisons with and without this adjustment for both color and surface codes (Fig.~\ref{fig:thresholds_cmp}) further confirm the effectiveness of this weighting strategy. Furthermore, the improvement observed in the surface code threshold implies that this adjustment can also adapt to enhance the correlated decoder for surface codes. 

The remaining boundary failures after adjustment are characterized by adjacent diagonal error pairs and quantified by $N_{\textrm{b}}$ in Eq.~\ref{eq:N_extra_color_correlated}. Table~\ref{tab:enumerating_error} also provides theoretical estimates for these residual errors across $d=6$ to 14, which align closely with the simulated values $N'$. Representative examples of the corresponding boundary failure configurations are shown in Fig.~\ref{fig:remaining_2D_error}.

\begin{table}[hb]
    \centering
    \begin{tabular}{|c|c|c|c|c|}
    \hline
    $d$ & Pattern & $N$& $N'$ & $N_\textrm{b}(i,j)$ \\
    \hline
    6 & \{D,S,N\} & 6 & 0 & 0\\
    \hline
    \multirow{2}{*}{8} & \{D,S,S,N\} & 72 & 0 & 0\\
     & \{D,D,N,N\} & 4 & 4 & 6  \\
    \hline
    \multirow{3}{*}{10} & \{D,D,S,N,N\} & 88 & 88 & 96 \\
     & \{D,E,S,N,N\} & 92 & 0 & 0\\
     & \{D,S,S,S,N\} & 584 & 0 & 0\\
    \hline
    \multirow{4}{*}{12}  & \{D,D,E,N,N,N\} & 64 & 64 & 80 \\
                         & \{D,D,S,S,N,N\} &1036& 896 & 960  \\          
                         & \{D,E,S,S,N,N\} &1676& 0 &  0 \\  
                         & \{D,S,S,S,S,N\} &3888& 0 &  0 \\          
    \hline
    \multirow{5}{*}{14}  & \{D,D,D,S,N,N,N\} & 288 & 0 & 0\\
                         & \{D,D,E,S,N,N,N\} &1888& 1888 & 1920 \\
                         & \{D,E,E,S,N,N,N\} &864 & 0 &0 \\
                         & \{D,D,S,S,S,N,N\} &10040& 7360 & 7680  \\
                         & \{D,E,S,S,S,N,N\} &19472&0&0 \\
                         & \{D,S,S,S,S,S,N\} &23184&0&0 \\
    \hline
    \end{tabular}
    \caption{Enumeration for extra boundary errors in a single row from $d=6$ to 14. $N$ and $N'$ represent simulated failure counts with $w_b=1$ and $w_b=0.999$. $N_\textrm{b}(i,j)$ is the analytical estimate of $N'$ from Eq.~\ref{eq:N_extra_color_correlated} for specific values of $i$ and $j$ within a boundary row. Minor deviations between simulation and theory may arise from decoder certainty and preference.} 
    \label{tab:enumerating_error}
\end{table}

\end{document}